\documentclass[journal=jacsat,manuscript=article]{achemso}

\usepackage[version=3]{mhchem} 
\usepackage{amssymb}
\usepackage{array}
\usepackage{makecell} 
\usepackage{xcolor}
\usepackage{xr}
\usepackage[hidelinks]{hyperref}

\usepackage[utf8]{inputenc} 
\usepackage[T1]{fontenc} 
\DeclareUnicodeCharacter{2212}{-} 
\DeclareUnicodeCharacter{2009}{\,} 

\makeatletter
\newcommand*{\addFileDependency}[1]{
\typeout{(#1)}
\@addtofilelist{#1}
\IfFileExists{#1}{}{\typeout{No file #1.}}
}\makeatother

\newcolumntype{P}[1]{>{\centering\arraybackslash}p{#1}}

\author{Hongyu Gao}
\affiliation[Saarland University]{Department of Materials Science \& Engineering, Saarland University, Campus C6.3, 66123 Saarbr\"ucken, Germany}
\email{hongyu.gao@uni-saarland.de}
\phone{+49 681-302-57458}
\author{Minghe Zhu}
\affiliation[Saarland University]{Department of Materials Science \& Engineering, Saarland University, Campus C6.3, 66123 Saarbr\"ucken, Germany}
\author{Jia Ma}
\affiliation[Changsha University of Science and Technology]{School of Civil and Environmental Engineering, Changsha University of Science and Technology, Changsha 410114, PR China}
\alsoaffiliation[Saarland University]{Department of Materials Science \& Engineering, Saarland University, Campus C6.3, 66123 Saarbr\"ucken, Germany}
\author{Marc Honecker}
\affiliation[Saarland University]{Department of Materials Science \& Engineering, Saarland University, Campus C6.3, 66123 Saarbr\"ucken, Germany}
\author{Kexian Li}
\affiliation[Changsha University of Science and Technology]{School of Civil and Environmental Engineering, Changsha University of Science and Technology, Changsha 410114, PR China}

\title[]{Probing Rate-Dependent Liquid Shear Viscosity Using Combined Machine Learning and Non-Equilibrium Molecular Dynamics}

\abbreviations{IR,NMR,UV}

\begin{document}








\begin{abstract}

Accurately measuring liquid dynamic viscosity across a wide range of shear rates, from the linear-response to shear-thinning regimes, presents significant experimental challenges due to limitations in resolving high shear rates and controlling thermal effects.
In this study, we integrated machine learning (ML) with non-equilibrium molecular dynamics (NEMD) simulations to address these challenges. 
A supervised artificial neural network (ANN) model was developed to predict viscosity as a function of shear rate, normal pressure, and temperature, effectively capturing the complex interplay among these variables.
The model reveals distinct trends in shear viscosity, characterized by the shear-thinning exponent, and highlights non-monotonic behavior in the radius of gyration components, reflecting molecular morphological changes driven by rate-dependent volume expansion.
Notably, temperature effects diminish at higher shear rates, where molecular alignment and spacing dominate the response to shear. 
By implementing the `\texttt{fix~npt/sllod}' command in LAMMPS, we achieve precise constant-pressure control in NEMD simulations, ensuring accurate representation of system dynamics. 
This study demonstrates the potential of ML-enhanced NEMD for efficient and accurate viscosity prediction, providing a robust framework for future research in complex fluid dynamics and material design.

\end{abstract}

\section{Introduction}

Dynamic viscosity ($\eta$), defined as the ratio of shear stress ($\tau$) to shear rate ($\dot\gamma$):
$\eta=\tau / \dot\gamma$,
quantifies a fluid's resistance to deformation under shear.
In the linear-response regime at low $\dot\gamma$, shear stress is proportional to shear rate ($\tau\propto\dot\gamma$), resulting in constant viscosity known as Newtonian viscosity ($\eta_0$). 
Beyond a critical shear rate ($\dot\gamma_0$), many liquids, including colloidal suspensions and polymer melts, exhibit shear-thinning behavior~\cite{Bair2002PRL}, where viscosity decreases with increasing $\dot\gamma$.
The nature of this transition---whether it follows a logarithmic dependency (Eyring theory~\cite{Eyring1936JCP}) or a power-law relationship (Carreau model~\cite{Carreau1972TSR})---remains a subject of debate~\cite{Spikes2014TL, Jadhao2017PNAS, Bair2015TL, Bair2017PNAS}.
This transition from Newtonian (or Stokesian~\cite{Mueser2020L}) to shear-thinning behavior is attributed to flow-induced molecular and microstructural  changes~\cite{Lemarchand2015JCP, Jadhao2019TL, Gao2024TL}, which reduce momentum transfer and lower energy barriers as molecules slide past one another.

Probing viscosity across a wide range of $\dot\gamma$ presents significant challenges in both experiments and simulations.
Experimental techniques, such as tribometry and viscometry~\cite{Spikes2014TL}, struggle at high $\dot\gamma$ due to thermal heating effects~\cite{Archard1959W}, which cause volume expansion~\cite{Daivis1994JCP} and potential underestimation of $\eta$.
Non-equilibrium molecular dynamics (NEMD) simulations~\cite{Gao2024TL} offer precise control over parameters like temperature ($T$), normal pressure ($P_{zz}$), and shear rate, providing an alternative for studying viscosity.
However, NEMD simulations face limitations, include long runtimes required to achieve sufficient signal-to-noise ratios at low $\dot\gamma$, thereby making it difficult to accurately predict Newtonian viscosity for highly viscous systems. 
Additionally, the accuracy of NEMD predictions depends on factors such as the choice of molecular model (explicit or coarse-grained), force field reliability~\cite{Ewen2016M}, and thermostating methods~\cite{Delhommelle2001JCP}.
Reconciling experimental and simulation results across overlapping $\dot\gamma$ regimes remains challenging, particularly for liquids with complex molecular structures or under extreme conditions.

Data-driven machine learning (ML) approaches have emerged as powerful tools for predicting material properties and accelerating material design~\cite{Ramprasad2017CM, Bartok2017SA, Dajnowicz2022JPCB}. 
For instance, physics-informed quantitative structure-property relationship (QSPR) models~\cite{Suzuki2001JCICS, Kauffman2001JCICS}, utilizing descriptor-based or graph-based neural networks, have successfully predicted kinematic viscosity, reducing experimental costs.
When integrated with molecular dynamics (MD) simulations~\cite{Yasuda2023AMI, Panwar2024JCIM}, these models achieved enhanced predictive accuracy.
However, ML models for dynamic viscosity remain scarce~\cite{Kadupitiya2021TL, Yasuda2023AMI, Sterr2024CM} due to limited datasets encompassing diverse liquid types and conditions, as well as challenges in extrapolating beyond training data.
Experimental acquisition of domain-specific knowledge is often resource-intensive, highlighting the importance of computational simulations with physical interpretability.
Dynamic viscosity, $\eta(\dot\gamma)$, under varying $T$ and $P_{zz}$ conditions often follows time-temperature-pressure superposition (TTPS)~\cite{Bird1987, Bair2014T}, which enables data normalization onto a master curve~\cite{Bair2002PRL, Jadhao2017PNAS}.
However, implementing TTPS requires prior knowledge of $\eta_0$ and $\dot\gamma_0$, which are themselves often targets of prediction, adding complexity to the process.

In this work, we developed a ML model to predict the rate-dependent dynamic viscosity of $n$-hexadecane, a representative liquid, by integrating all-atom NEMD simulations.
Our model, based on an artificial neural network (ANN), is trained on NEMD simulation data and validated for bulk-phase liquids, distinguishing it from studies of confined fluids in nanoscale slits~\cite{Gao2020JCIS, Gao2022L}, where oscillatory density profiles typically arise. 
By emphasizing the importance of constant-pressure control, we systematically explored the dependences of $\eta$ on $T$ and $P_{zz}$ in the shear-thinning regime, capturing the interplay between molecular dynamics and macroscopic flow behavior.
Despite challenges posed by limited training data, our approach demonstrates robust predictive accuracy and computational efficiency, providing a scalable framework for dynamic viscosity prediction.
This work is expected to advance the understanding of fluids shear-thinning behavior and to establish a foundation for extending ML-driven approaches to more intricate systems and extreme conditions.
%

\section{Methodology}

\subsection{Molecular Dynamics}

Both non-equilibrium molecular dynamics (NEMD) and equilibrium molecular dynamics (EMD) simulations were performed using the open-source code -- LAMMPS~\cite{Plimpton1995JCP}.
The simulation cell consisted of 100 $n$-hexadecane molecules, a system size optimized to minimize finite-size effects.
$n$-hexadecane was selected as the model system due to prior experience~\cite{Gao2024TL, Oo2024NL} and its computational efficiency.
An all-atom representation was employed to ensure accurate shear stress calculations, as preliminary tests using the united-atom TraPPE force field~\cite{Martin1998JPCB} revealed a significant underestimation of shear stress by at least 20\%.
Bonded interactions (bond stretching, angle bending, and torsion) and non-bonded interactions (van der Waals and Coulombic forces) were described using the L-OPLS~\cite{Price2001JCC, Siu2012JCTC} force field, which is specifically optimized to predict phase transition temperatures of long-chain alkanes.
EMD simulations were primarily used to validate Newtonian viscosity via the Green-Kubo method~\cite{Green1954JCP, Kubo1957JPSJ}, with further methodological details provided in the Supporting Information (SI).
Unless otherwise noted, references to ``simulation results'' pertain to those obtained from NEMD simulations.

Shear viscosity was calculated as the ratio of shear stress ($\tau$) to shear rate ($\dot\gamma$) under linear planar (Couette) flow conditions, where the velocity gradient perpendicular to the shear plane remains as constant.
This approach avoids the molecular layering effects associated with solid wall-induced boundary-driven shear~\cite{Gao2020JCIS}, thereby providing a true representation of bulk liquid viscosity.
Lees-Edwards~\cite{Lees1921JPCSSP} equivalent periodic boundary conditions (PBC) were applied to remap atom positions and velocities crossing the simulation boundaries, while the standard SLLOD algorithm~\cite{Evans1984PRA} was employed to model both conservative and dissipative forces.
Simulations were performed under the $nPT$ ensemble, regulated by the Nos\'e-Hoover~\cite{Nose1984MP, Hoover1985PRA} thermostat and barostat to maintain system temperature and pressure.
The choice of $nPT$/SLLOD over $nVT$/SLLOD is justified in the Results section.
Shear rate, temperature, and normal pressure ($P_{zz}$) were varied across ranges of $10^8\sim 10^{12}$ 1/s, $300\sim 400$ K, and $100\sim300$ MPa, respectively, with a simulation timestep of 1 fs.
Error bars were calculated from uncorrelated data achieved, with the dump frequency optimized via time-autocorrelation function ($t$-ACF) analysis.

\subsection{Machine Learning}

A machine learning (ML) model was developed using an artificial neural network (ANN) regressor trained on data derived from non-equilibrium molecular dynamics (NEMD) simulations.
The input features included applied conditions: shear rate ($\dot\gamma$), temperature ($T$), and normal pressure ($P_{zz}$), with the target output being liquid shear viscosity ($\eta$).
The ANN was selected as the final regression algorithm due to its superior performance compared to other tested models, including linear regression, random forest, extra trees, gradient boosting, support vector machines, and $k$-nearest neighbors.
Ensemble methods such as voting and stacking regressors were also evaluated but did not outperform the ANN. 
To capture rate-dependent structural and thermodynamic factors, additional variables such as density ($\rho$) and radius of gyration components ($R_g^x$, $R_g^y$, and $R_g^z$) were incorporated into the input features. 
However, their contributions to prediction accuracy were found to be marginal, suggesting that the primary input features ($\dot\gamma, T, P_{zz}$) were sufficient for accurate viscosity prediction.

The dataset was randomly split into 80\% for training and 20\% for validation to ensure unbiased evaluation of model performance.
To address limited data dimensionality, five replicate results from parallel simulation runs were included for each unique set of conditions ($\Dot{\gamma}, T, P_{zz}$), enhancing the model's ability to capture variability and improve generalization.
Logarithmic transformations were applied to $\dot\gamma$ and $\eta$ to account for their broad value ranges and inherent non-linear relationships.
The models were implemented using the scikit-learn and Keras libraries. 
The ANN architecture consisted of a feed-forward network with two hidden layers (64 and 32 neurons, respectively), utilizing the ReLU activation function and the Adam optimizer for efficient training convergence.
Hyperparameter optimization was conducted via randomized search combined with 5-fold cross-validation to ensure robust parameter selection.
The validity of the ML model was benchmarked against the NEMD results under identical conditions.
Model performance was comprehensively evaluated using metrics such as mean squared error (MSE), mean absolute error (MAE), $R$-squared ($R^2$), and root mean squared error (RMSE), proving a thorough evaluation across the varying scales of $\eta$.
Further validation, including fitting to the Carreau-Yasuda model, is detailed in the Results section.

\section{Results and Discussion}

\subsection{Running NEMD under an $nPT$ ensemble}

\begin{figure}[ht]
\centering
\includegraphics[width=0.42\textwidth]{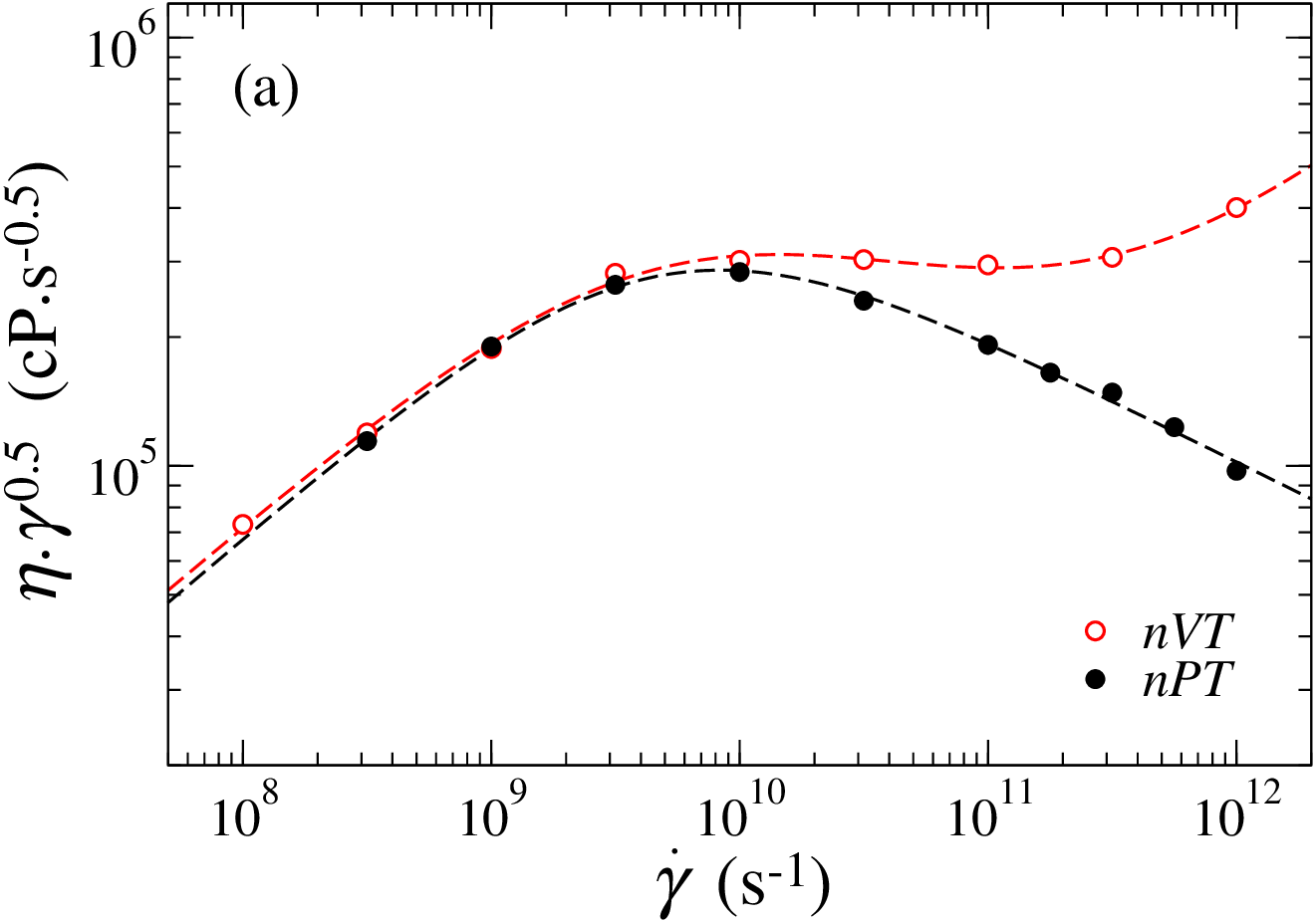}~~~~
\includegraphics[width=0.42\textwidth]{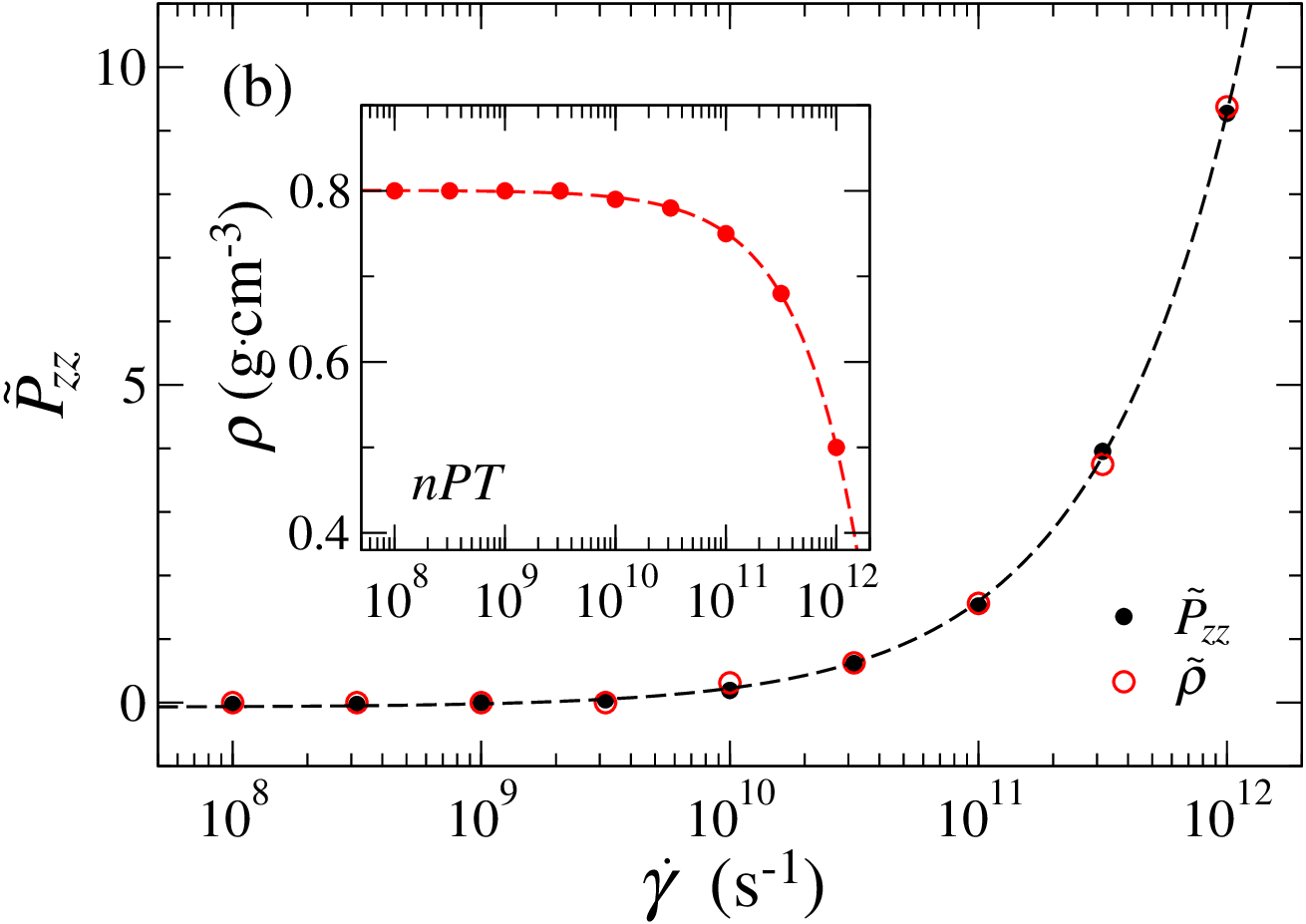}~~~~
\caption{
(a) Intermediate scaling plot of shear viscosity ($\eta$) as a function of shear rate ($\dot\gamma$) from constant-density ($\rho$) and constant-normal-pressure ($P_{zz}=100$ MPa) simulations at temperature $T=300$ K.
(b) Collapse of the normalized density ($\tilde{\rho}$) onto the normalized $\tilde{P}_{zz}$ curve as a function of $\dot\gamma$ with a proportionality constant $\alpha$=2.5.
The inset shows the variation of $\rho$ with $\dot\gamma$ under const.-$P_{zz}$ control. 
}
\label{fig:nvt_npt}
\end{figure}

In the shear-thinning regime, molecules moving rapidly along the streaming direction lack sufficient time to relax and dissipate energy arising from intense atomic collisions and internal friction.
Although excess heat can be removed via velocity rescaling, molecular morphological changes under constant normal pressure ($P_{zz}$) result in system volume expansion and a corresponding drop in density ($\rho$), as illustrated in the inset of Fig.~\ref{fig:nvt_npt}b.
This phenomenon leads to an overestimation of $\tau$ at high $\dot\gamma$ under constant density (const.-$\rho$) control due to elevated hydrostatic pressure.
To address this, we modified the {LAMMPS} source code to implement a new command, `\texttt{fix~npt/sllod}', which enables barostat control exclusively in the normal direction while incorporating the {SLLOD} algorithm to handle in-plane cell deformation.
By regulating only $P_{zz}$, this approach avoids conflicts with imposed in-plane domain deformation, where atom positions and velocities are remapped based on the Lees-Edwards periodic boundary conditions (PBC)~\cite{Todd2017}.
This modification enhances physical fidelity, maintains system stability during non-equilibrium dynamics, and significantly improves modeling efficiency compared to pressure interpolation methods (see SI for details).
The modified C++ source code is provided in the supplementary materials.

Predictions from const.-$\rho$ ($nVT$ control) and const.-$P_{zz}$ ($nPT$ control) are compared through intermediate scaling plots~\cite{Gao2024TL}, where data are fitted using the Carreau-Yasuda (CY) model~\cite{Carreau1972TSR, Yasuda1981RA}:
\begin{equation}
\eta(\dot\gamma)=\eta_{\infty}+(\eta_{0}-\eta_{\infty})[1+(\frac{\dot\gamma}{\dot\gamma_{0}})^a]^{\frac{n-1}{a}},
\end{equation}
where $\eta_{\infty}$ represents the second Newtonian viscosity, and $a$ and $n$ are fitting parameters defining the crossover curvature and shear-thinning exponent, respectively.
As shown in Fig.~\ref{fig:nvt_npt}a, predictions in the linear response regime are comparable; however, differences grow beyond the crossover and become pronounced at higher $\dot\gamma$.
Consistent with the findings of Daivis and Evans~\cite{Daivis1994JCP}, $\eta_{\infty}$ is only necessary for const.-$\rho$ predictions, serving as an artificial adjustment.
The CY model demonstrates superior performance over the Eyring theory~\cite{Eyring1936JCP} (as detailed in Ref.~\cite{Gao2024TL}), achieving a lower normalized logarithmic relative standard deviation, even when accounting for the parameter count ($N_{\rm CY}$=4 versus $N_{\rm Eyring}$=2).

In the Newtonian regime, the diagonal pressure tensor components ($P_{ii}$, $i=x,y,z$) remain comparable to the isotropic hydrostatic pressure observed in bulk equilibrium in the absence of shear.
However, as the strain rate increases and the system enters the shear-thinning regime, these pressures diverge, with $P_{zz}$ exhibiting an exponential rise under const.-$\rho$ conditions.
Consequently, the relationship between pressure and density cannot be adequately described by traditional equations of state (EOS), such as the Tait~\cite{Tait1888} or Murnaghan~\cite{Murnaghan1944PNAS} equations, due to shear-induced anisotropy~\cite{Gao2024TL}.
As shown in Fig.~\ref{fig:nvt_npt}b, the variation in normal pressure ($\tilde P_{zz}$) under const.-$\rho$ is proportional to the variation in density ($\tilde\rho$) under const.-$P_{zz}$:
\begin{equation}
\tilde P_{zz} = \alpha \tilde\rho,
\end{equation}
where $\tilde P_{zz}=P_{zz}^{nVT}/{P_{zz}^{nPT}}-1$, $\tilde\rho=1-\rho^{nPT}/\rho^{nVT}$, and $\alpha$ is a proportionality constant.
Nevertheless, the crossover observed in the $\dot\gamma-\rho$ (or $\dot\gamma-P_{zz}$) curves does not align with those in the $\dot\gamma-\eta$ plots, indicating that changes in density (or normal pressure) along do not fully explain the shear-thinning behavior.
This highlights the need for further investigation into molecular morphology and microstructural changes under shear flow to better understand variations in shear stress.

\subsection{ML-Predicted Shear Viscosity}

\begin{figure}[ht]
\centering
\includegraphics[width=0.45\textwidth]{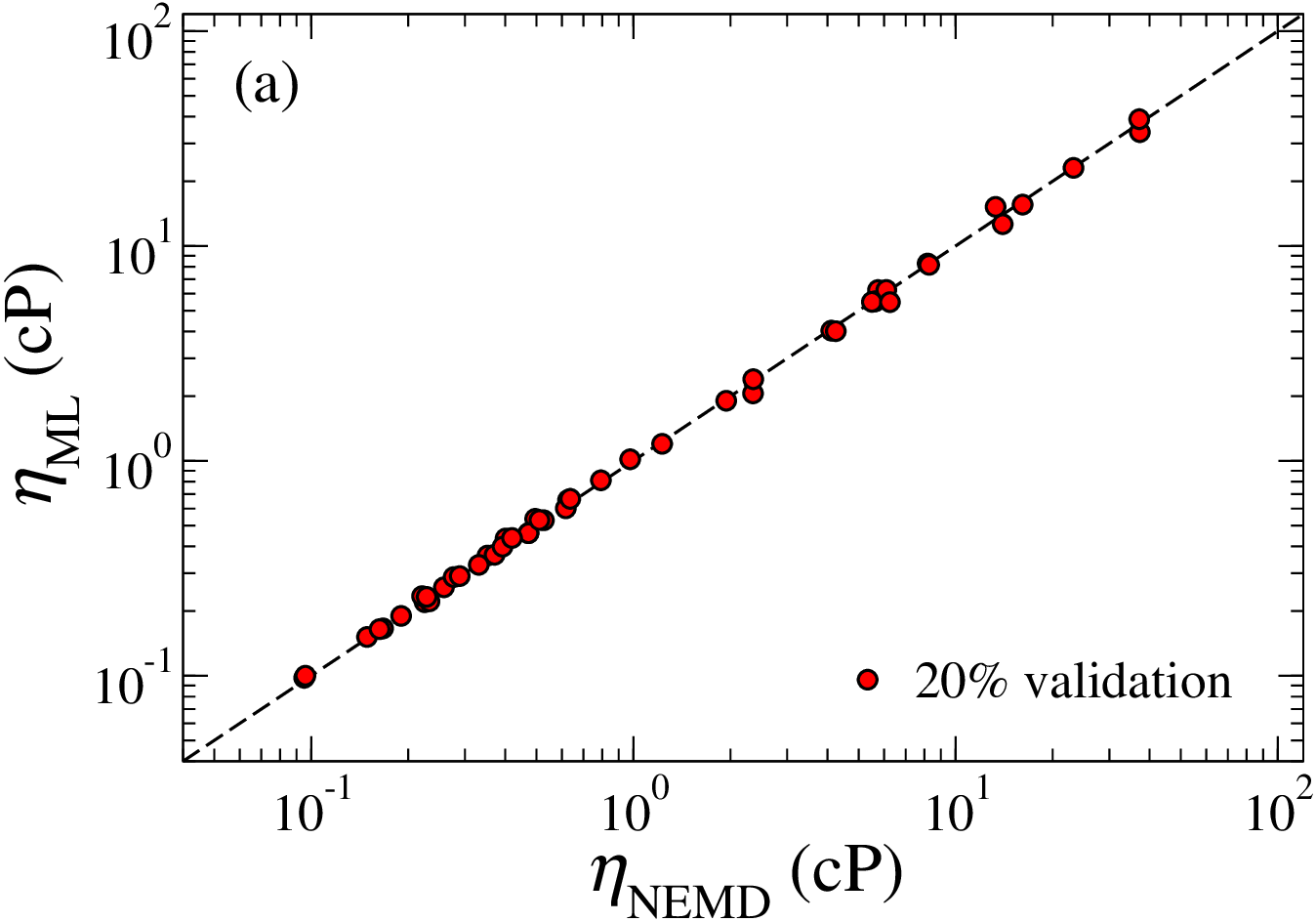}~~~~
\includegraphics[width=0.45\textwidth]{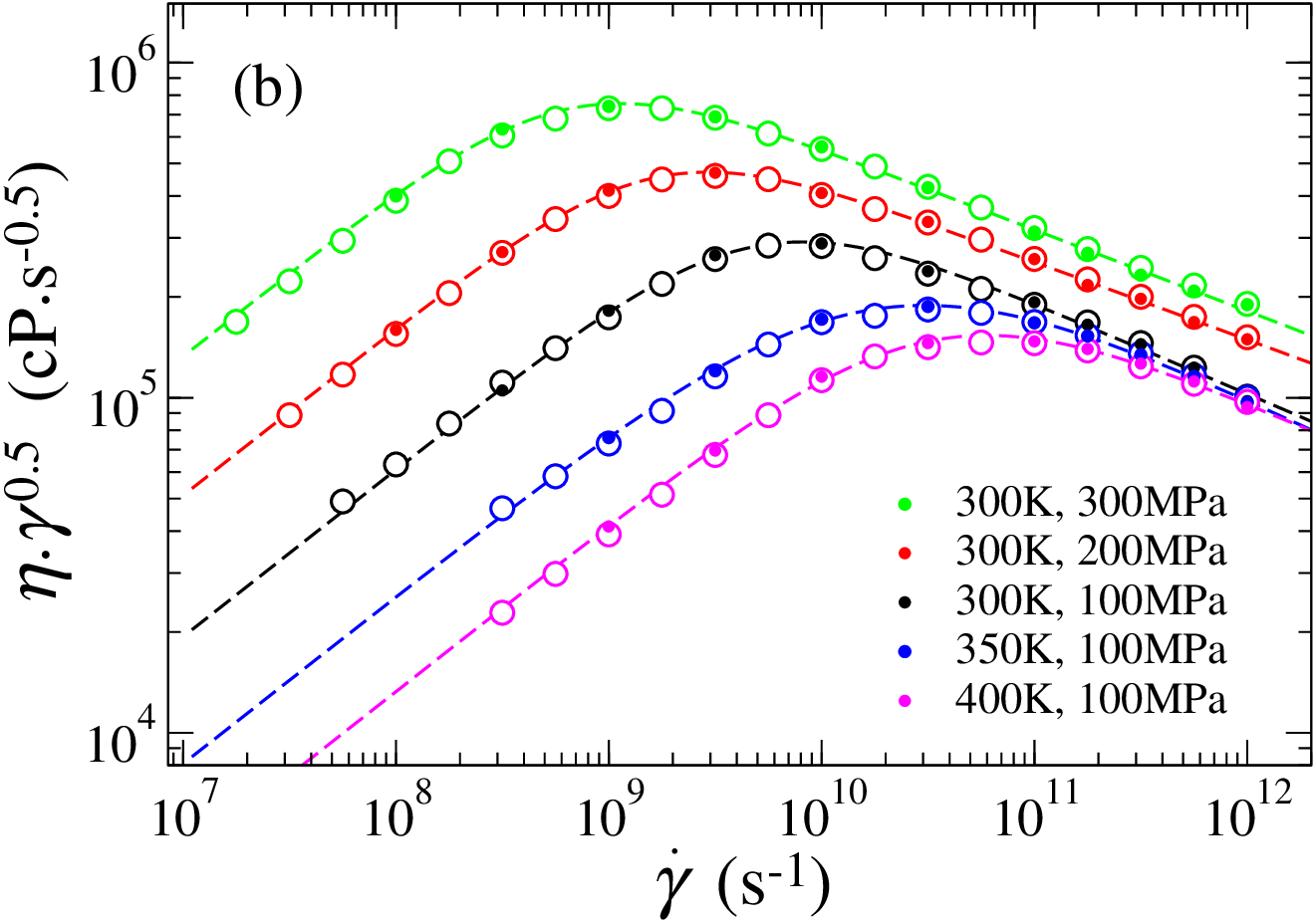}
\caption{
Comparison of shear viscosity ($\eta$) predicted by NEMD modeling and machine learning (ML) for (a) the 20\% validation dataset and (b) additional untrained data at temperatures $T\in$ \{300, 350, 400 K\} and normal pressures $P_{zz}\in$ \{100, 200, 300 MPa\}.
In (b), solid symbols denote NEMD predictions, hollow symbols represent ML predictions, and dashed lines indicate Carreau-Yasuda (CY) fits to the NEMD data.
}
\label{fig:ML_viscosity}
\end{figure}

The NEMD dataset, though limited in size, provides high-quality data derived from simulations averaged over sufficiently long times with a high signal-to-noise ratio.
Using three input features, i.e., $\dot\gamma$, $T$, and $P_{zz}$, the ANN model demonstrates exceptional performance in reproducing the NEMD-predicted $\eta$ across a wide range of $\dot\gamma$ and moderate variations in $T$ and $P_{zz}$.
The model achieved an impressive MSE of 4.8$\times 10^{-4}$, MAE of 1.6$\times 10^{-2}$, and an R$^2$ score of 0.9991 on the training set, using a fixed random seed of 42.
Training employed the Adam optimizer with a learning rate of 0.001, a batch size of 8, and 2000 epochs (these batch size and epoch values were selected as the best-performing configuration from a series of tests via hyperparamter tuning using a pipeline).
Validation loss closely tracked the training loss, with both decaying approximately as 1/$x$ with epochs, indicating effective and consistent learning.
The small batch size may have introduced noise into the gradient updates, potentially helping the model explore the loss landscape more effectively and improving generalization performance. 
Alternative optimization algorithms, such as root mean square propagation (RMSProp) and Levenberg-Marquardt~\cite{Marquardt1963JSIAM} (LM), produced comparable results when tuned to their respective optimal hyperparameters.
Validation results, compared in Fig.~\ref{fig:ML_viscosity}a, reveal close agreement between ML-predicted and NEMD viscosity, underscoring the model's reliability in capturing the target behavior.

To further evaluate the model, predictions were performed under ($T, P_{zz}$) conditions consistent with NEMD simulations, enabling direct comparisons through CY fitting.
As shown in Fig.~\ref{fig:ML_viscosity}b, the ANN model predictions (hollow symbols) not only reproduced the NEMD results (solid symbols) but also extended seamlessly along the CY fitting curves (dashed lines), achieving an $R^2$ of 0.99.
Notably, the NEMD data in Fig.~\ref{fig:ML_viscosity}b were not part of the training set, highlighting the model's ability to generalize beyond the training data.
We would like to emphasize that the predictive accuracy of the model is largely influenced by the quality of the input data, particularly in the crossover region, which plays a crucial role in determining the zero-shear viscosity ($\eta_0$).
Compared to other regression algorithms tested, the ANN model exhibits superior interpolation capabilities.
However, accurate extrapolation to the Newtonian regime depends heavily on the quality and availability of the training input data, highlighting the need for further optimization in scenarios requiring precise extrapolation, particularly in high-viscosity systems.

\begin{figure}[ht]
\centering
\includegraphics[width=0.45\textwidth]{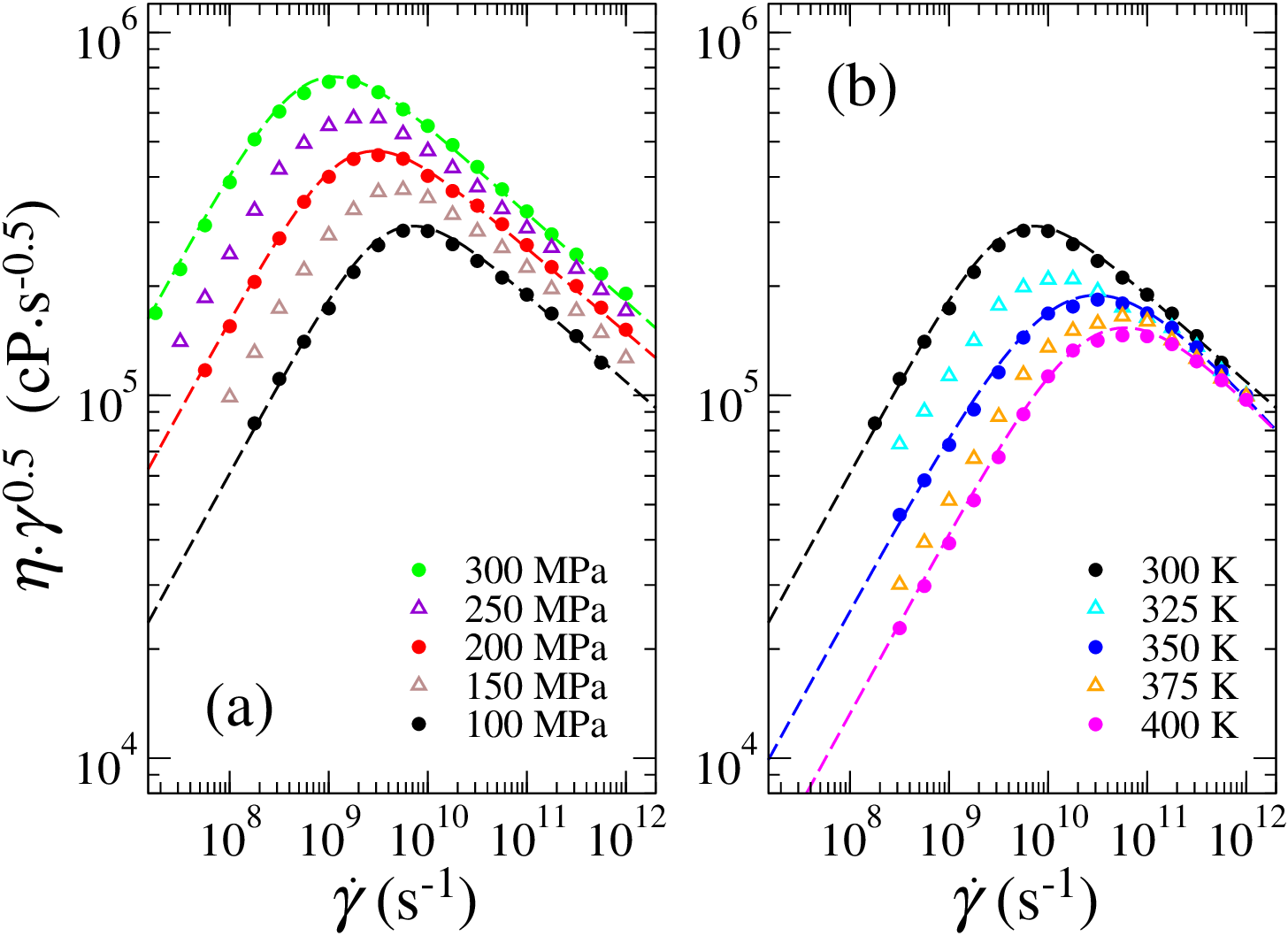}~~
\includegraphics[width=0.48\textwidth]{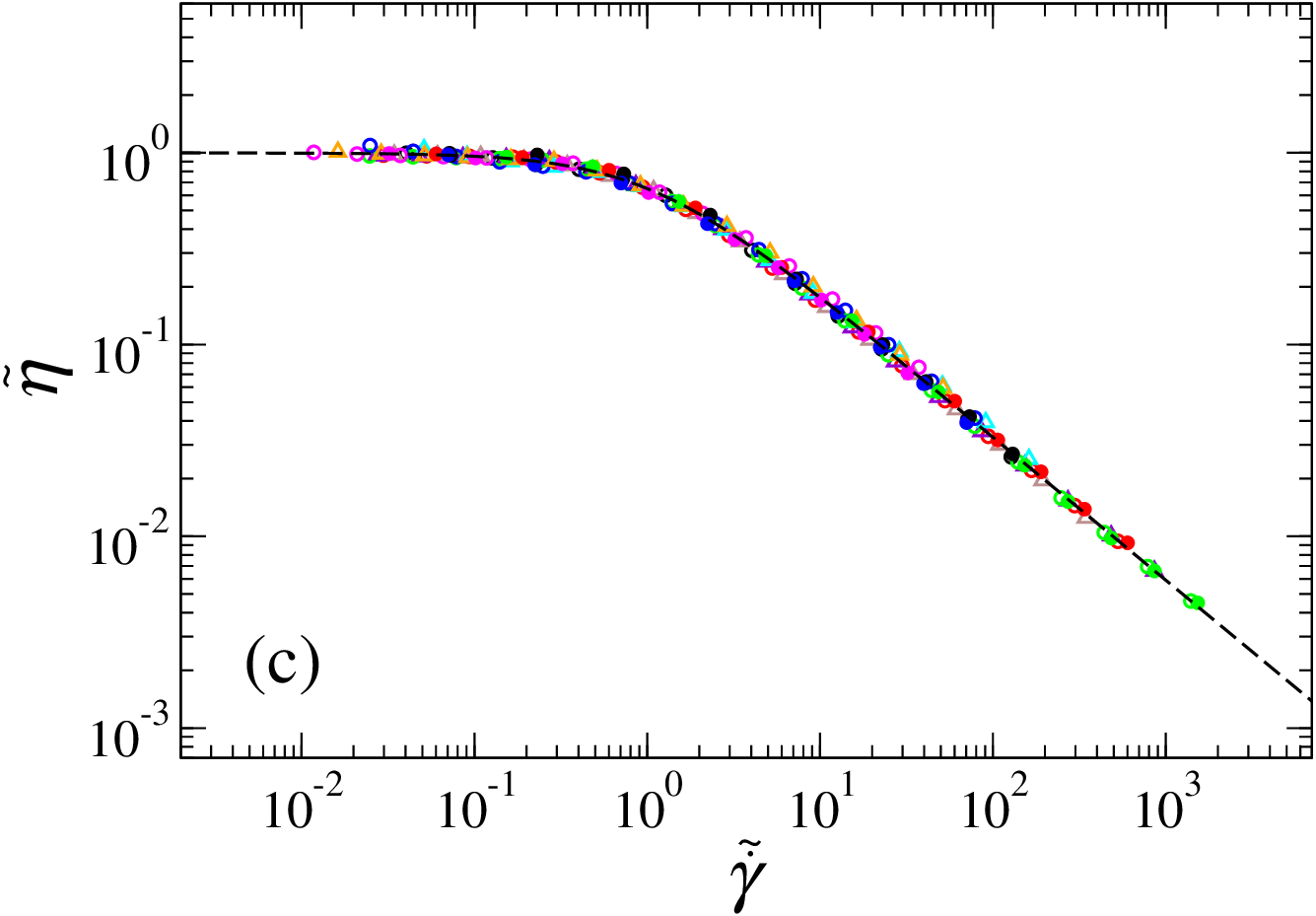}
\caption{
Intermediate scaling plot of machine learing (ML) predictions under interpolated conditions for (a) normal pressure ($P_{zz}$) and (b) temperature ($T$), represented by hollow triangles.
Dashed lines in (a) and (b) correspond to Carreau-Yasuda (CY) fits to the NEMD data for reference.
(c) Collapse of all ML-predicted results (hollow symbols) and NEMD data (solid symbols) onto a master curve, expressed in terms of normalized viscosity ($\tilde\eta$) and normalized shear rate ($\tilde{\dot\gamma}$). 
The dashed line represents CY fitting with $\tilde\eta_0=1$ and $\tilde{\dot\gamma}_0=1$. 
}
\label{fig:ML_validation_master}
\end{figure}

Predictions were extended to conditions beyond those covered by NEMD results, with CY fitting serving as a reference to rationalize prediction trends.
Although indirect, the CY fitting method reliably captured the rate-dependent $\eta$, particularly the shear-thinning behavior at high $\dot\gamma$ under specific $T$ and $P_{zz}$ conditions~\cite{Gao2024TL}.
To further validate the representation of CY fittings within the linear-response regime, particularly when the NEMD data are sparse near the crossover, zero-shear $\eta_0$ derived from EMD simulations was used as a benchmark.
As detailed in the SI, $\eta_0$ from EMD simulations closely matched those obtained from CY fitting, affirming the latter's validity as a reference.
Leveraging these benchmarks, additional interpolated ML predictions were generated for new ($\eta, T, P_{zz}$) conditions (Figures~\ref{fig:ML_validation_master}a and~\ref{fig:ML_validation_master}b, solid triangles), which adhere well to expected trends based on uniform intervals of condition variations.
Beyond visual inspection, ML-predicted data effectively collapse onto a master curve (Fig.~\ref{fig:ML_validation_master}c) when normalized as
$\tilde\eta = {\eta} / {\eta_0}$ and
$\tilde{\dot\gamma} = {\dot\gamma} / {\dot\gamma_0}$,
where $\eta_0$ and $\dot\gamma_0$ were obtained from respective CY fittings, further demonstrating the consistency and reliability of the predictions.

Our supervised ANN model demonstrated exceptional performance in directly mapping physical parameters ($\dot\gamma$, $T$ and $P_{zz}$) to viscosity--the target outputs, providing significant advantages in scalability and efficiency.
Unlike unsupervised local dynamics ensemble (LDE)-based methods, this direct prediction (DP) approach requires less memory, eliminates molecular-type-specific adjustments, and simplifies implementation.
While computationally efficient, the DP approach relies heavily on the quality and diversity of the training dataset, posing potential challenges when extrapolating to conditions or molecular systems beyond the training range.
The model's predictions are grounded in well-defined physical parameters ($\dot\gamma,~T,~{\rm and}~P_{zz}$), making its behavior more transparent and easier to validate against known physical principles. 
%
%
Per our tests, incorporating additional input features, such as density ($\rho$) and radius of gyration ($R_g$), did not significantly improve prediction accuracy.
However, these parameters could be valuable as output features, providing insights into molecular dynamics and temporal trends.

\subsection{Roles of $T$ and $P_{zz}$}
 
\begin{figure}[ht]
\centering
\includegraphics[width=0.5\textwidth]{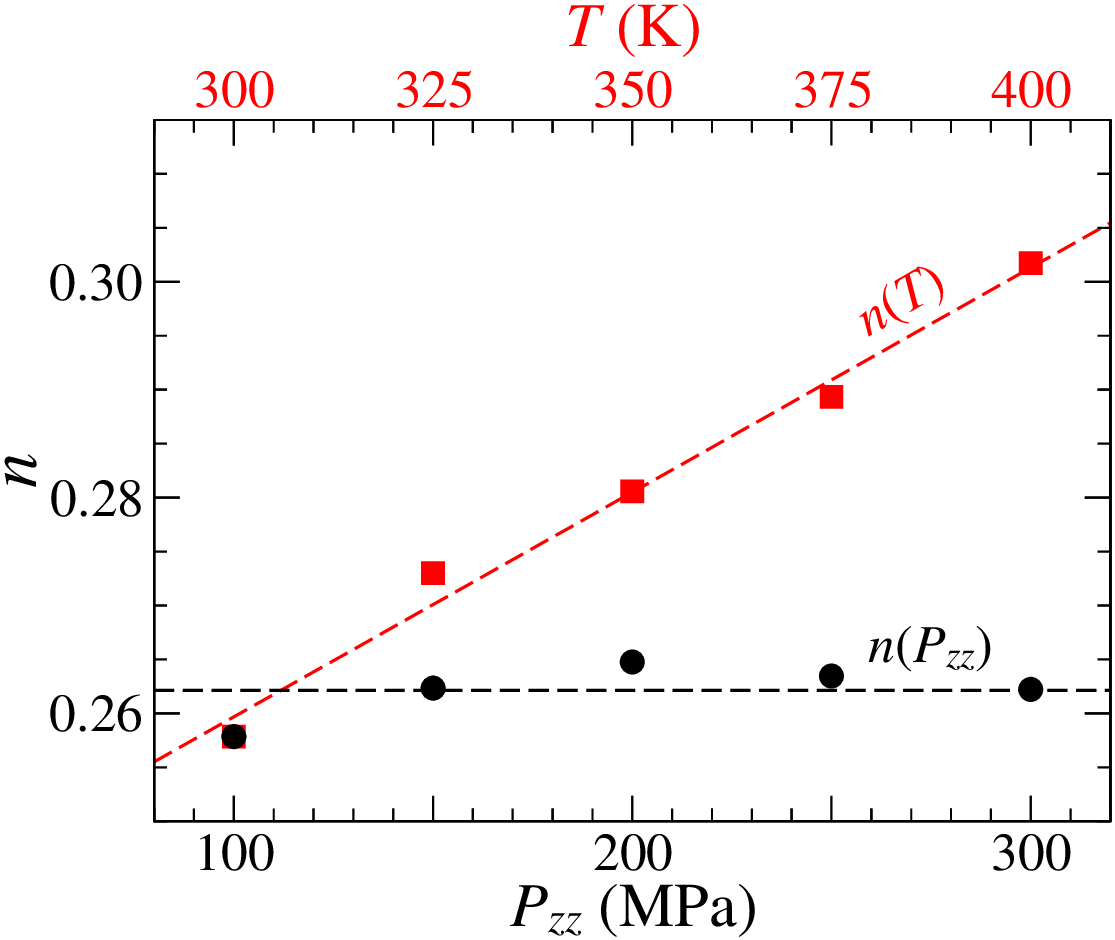}
\caption{
Variations of the shear-thinning exponent ($n$) obtained from Carreau-Yasuda (CY) fitting as functions of temperature ($T$, red) and normal pressure ($P_{zz}$, black).
}
\label{fig:CY_exponent}
\end{figure}

The dependences of equilibrium viscosity $\eta_0$ in the linear-response regime on $T$ and $P_{zz}$ have been explored in our previous work~\cite{Gao2024TL} for the same linear alkane system.
Notably, as $T$ decreases, a transition from non-Arrhenius to Arrhenius behavior is observed, corresponding to a fragile-to-strong transition.
Meanwhile, the dependence of $\eta_0$ on $P_{zz}$ follows a generalized hybrid function, where a power-low term dominates at low-to-negative $P_{zz}$, while an exponential term governs at moderate-to-high $P_{zz}$.
However, in the shear-thinning regime at high $\dot\gamma$, the dependences of $\eta (\dot\gamma)$ on $T$ and $P_{zz}$ exhibit distinct trends: parallel for $T$ (Fig.~\ref{fig:ML_validation_master}a) and converging for $P_{zz}$ (Fig.~\ref{fig:ML_validation_master}b).
The rate at which $\eta(T)$ and $\eta(P_{zz})$ decrease with increasing $\dot\gamma$ is quantified by the shear-thinning exponent $n$, obtained from CY fitting.
As illustrated in Fig.~\ref{fig:CY_exponent}, $n$ remains nearly constant across different $P_{zz}$ but increases linearly with $T$.
This highlights distinct dynamical behaviors and suggests that assuming a constant $n$ or simply setting $n=$ 0.5 under various conditions can lead to significant fitting errors.

The trends observed in Fig.~\ref{fig:CY_exponent} contrast with those fitted using the Carreau model~\cite{Voeltzel2016TL}, where $n$ increases with $T$ but decreases with $P_{zz}$.
Generally, increasing temperature enhances molecular vibrational frequencies and induces volumetric expansion, weakening intermolecular forces and facilitating molecular alignment under shear.
In contrast, pressure primarily affects molecular proximity by increasing density and modulating the energy barrier during shear.
In the Newtonian regime, temperature effects typically dominate, as thermally induced volume change occurs more readily than those caused by external pressure due to the low compressibility of liquids.
While $T$ and $P_{zz}$ influence $\eta$ in a generally comparable manner~\cite{Gao2024TL}, both often exhibit exponential or stretched-exponential behavior.
At high $\dot\gamma$ in the shear-thinning regime, externally applied forces induce intense atomic collisions, leading to significant temperature rise.
However, in molecular simulations, temperature is regulated via velocity rescaling, which artificially alters system dynamics while ensuring that the intrinsic response to shear remains dominant.
Although removing thermal heating effects would clarify the individual contributions of $T$ and $P_{zz}$, this is practically impossible due to the limited thermal conductively of solid counterfaces.
This convergence persists in the shear-thinning regime as long as the system maintains consistent structural or phase responses, ensuring smooth variations in shear stress~\cite{Gao2024TL}.

\begin{figure}[ht]
\centering
\includegraphics[width=0.7\textwidth]{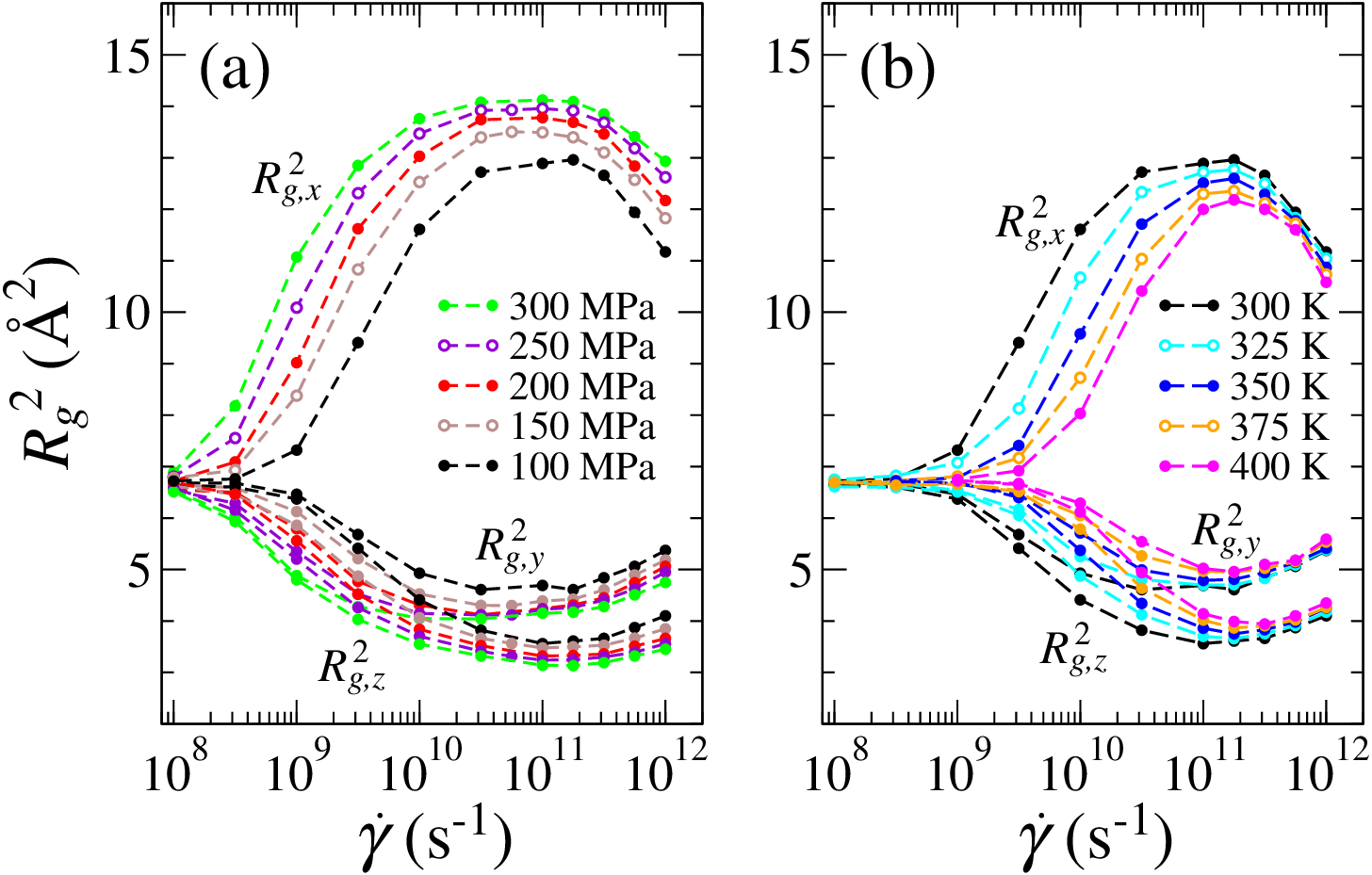}
\caption{
Radius of gyration components ($R_g^2$) as a function of shear rate ($\dot\gamma$) under varying (a) normal pressure ($P_{zz}$) at a constant temperature ($T$) of 300 K and (b) $T$ at a constant $P_{zz}$ of 100 MPa.
Solid symbols represent results from NEMD simulations, while hollow symbols denote predictions from the machine learning (ML) model.
}
\label{fig:rg}
\end{figure}

In the shear-thinning regime, as $\dot\gamma$ increases, liquid molecules tend to stretch in response to shear stresses, aligning their longitudinal direction parallel to the streaming direction. 
This morphological change facilitates sliding and can be characterized using the radius of gyration components:
\begin{equation}
R_g^2=
\langle
\frac{1}{M}\sum_i m_i(r_i-r_{\rm cm})^2
\rangle,
\end{equation}
where $M$ represents the total mass of a molecule, $r_i$ and $r_{\rm cm}$ denote the positions of the $i^{\rm th}$ monomer and the center of mass of the molecule, respectively.
Unlike the layering-like structure observed in liquids confined to nanometer-scale slits~\cite{Gao2020JCIS}, the spatially resolved density of bulk-phase liquid Couette flow remains constant even under relatively high normal pressure.
As shown in Fig.~\ref{fig:rg}, structural anisotropy emerges when the system enters the shear-thinning regime, varying non-monotonically with $\dot\gamma$. 
The alignment of molecules parallel to the streaming direction (indicated by higher $R_{g,x}^2$) reduces atomic collisions, leading to a lower $\tau(\dot\gamma)$ that deviates from the trend.
At high $\dot\gamma$, the rapid expansion of molecular spacing, evidenced by the significant density drop shown in Fig.~\ref{fig:nvt_npt}b, causes molecules to coil up again.
Further increasing $\dot\gamma$ can induce a liquid-to-gas phase transition, where the three $R_g^2$ components converge and become comparable.
While the dependences of $R_g(\dot\gamma)$ on $P_{zz}$ and $1/T$ appear similar at low $\dot\gamma$, the temperature effect becomes less significant when $R_g^2$ exceeds the extreme values as $\dot\gamma$ increases.

\section{Conclusions}

This work demonstrates the quantification of liquid shear viscosity ($\eta$) under coupled conditions of shear rate ($\dot\gamma$), temperature ($T$), and normal pressure ($P_{zz}$) using an integrated approach combining non-equilibrium molecular dynamics (NEMD) and machine learning (ML).
By employing a supervised artificial neural network (ANN), the model effectively and efficiently predicts $\eta(\dot\gamma,T,P_{zz})$ using NEMD-generated training data.
The ML approach provides a significant advantage by elucidating the complex relationships among $\dot\gamma$, $T$, and $P_{zz}$ in a reliable and computationally efficient manner, without the need for additional input features such as density and radius of gyration.
Future improvements could focus on enhancing the model's extrapolation capabilities to extended condition regimes, enabling predictions for more diverse and extreme scenarios.

We highlight the importance of constant-pressure control in NEMD simulations, achieved through the implementation of a modified LAMMPS command, `\texttt{fix~npt/sllod}'.
This ensures accurate representation of system dynamics under varying pressure conditions. 
Our results reveal distinct influences of $T$ and $P_{zz}$ on the shear-thinning regime: the shear-thinning exponent ($n$) from Carreau-Yasuda fitting remains constant across different $P_{zz}$ but increases linearly with $1/T$ within the studied range.
Additionally, the radius of gyration components exhibit non-monotonic trends as a function of $\dot\gamma$, reflecting molecular morphological changes driven by shear-induced alignment and volume expansion.
Notably, temperature effects become less significant at higher $\dot\gamma$, where shear-driven dynamics dominate the system's response.

%
%

\begin{acknowledgement}

We thank Martin M\"user for useful discussions.
This research was supported by the German Research Foundation (DFG) under grants number GA 3059/2-1 and GA 3059/4-1.

\end{acknowledgement}





\end{document}